# Cadmium Zinc Telluride (CZT) photon counting detector Characterisation for soft tissue imaging

*K. Hameed[1], Rafidah Zainon[1*] and Mahbubunnabi Tamal[2]*

[1]Department of Biomedical Imaging, Advanced Medical and Dental Institute, Universiti Sains Malaysia, SAINS@BERTAM, 13200 Kepala Batas, Pulau Pinang, Malaysia.

[2]Department of Biomedical Engineering, Imam Abdulrahman Bin Faisal University, PO Box 1982, Dammam 31441, Saudi Arabia.

*Corresponding:*
 *Mahbubunnabi Tamal
Email address: mtamal@iau.edu.sa; mtamal@yahoo.com

*Abstract*

The use of photon counting detection technology has resulted in significant X-ray imaging research interest in recent years. Computed Tomography (CT) scanners can benefit from photon-counting detectors, which are new technology with the potential to overcome key limitations of conventional CT detectors. Researchers are still studying the effectiveness and sensitivity of semiconductor detector materials in photon counting detectors for detecting soft tissue contrasts. This study aimed to characterize the performance of the Cadmium Zinc Telluride photon counting detector in identifying various tissues. An optimal frame rate per second (FPS) of CZT detector was evaluated by setting the X-ray tube voltage and current at 25 keV, 35 keV and 0.5 mA, 1.0 mA respectively by keeping the optimum FPS fixed, the detector energy thresholds were set in small steps from 15 keV to 35 keV and the Currents were set for X-ray tubes in ranges of 0.1 mA to 1.0 mA to find the relationship between voltage and current of the X-ray source and counts per second (CPS). The samples i.e., fat, liver, muscles, paraffin wax, and contrast media were stacked at six different thickness levels in a stair-step chamber made from Plexi-glass. X-ray transmission at six different thicknesses of tissue samples was also examined for five different energy (regions) thresholds (21 keV, 25 keV, 29 keV, 31 keV, and 45 keV) to determine the effect on count per second (CPS). In this study, 12 frames per second is found to be the optimum frame rate per second (FPS) based on the spectral response of an X-ray source and CPS has a linear relationship with X-ray tube current as well. It was also noted that A sample's thickness also affects its X-ray transmission at different energy thresholds.  A high sensitivity and linearity of the detectors make them suitable for use in both preclinical and medical applications.

## *1. Introduction*

An X-ray detector is a key component in X-ray-based medical imaging modalities because it affects the image quality. Conventional X-ray detectors consist of two main components such as a scintillator crystal, that converts incoming X-ray photons into visible light and photodiodes or charge couple devices, that convert visible light into electrical current, which is recorded by the data acquisition system.

Conventional x-ray detectors integrate the total electrical current produced in the radiation sensor and disregard the charge amplitude from individual photon detection events. The charge amplitude from each event is proportional to the photon's detected energy. During this integration, both the detector leakage current and charges resulting from X-ray detection are summed and measured, and provide no information about the energy of individual photons or the dependence of the attenuation coefficients in the object [1]. However, the integration of the data results in the loss of vital information carried by the energy of each incoming X-ray photon regarding the X-ray linear attenuation coefficient ($\mu$) of the object, which is related to its chemical structure and mass density. This type of detector is known as an energy integrating detector (EID) [2].

Photon counting detectors (PCDs), which contain room temperature semiconductors such as cadmium telluride (CdTe) [3] or cadmium zinc telluride (CZT) [4] are novel alternatives to EIDs. They overcome the limitations of EIDs by providing the ability to count photons based on their detected energies [5, 6]. PCDs allow photons to be grouped into multiple energy bins based on the energy threshold selected for each individual bin [7].

**Figure 1** shows graphical illustration of X-ray photon detection process and recording using Photon Counting Detector (PCD).

Recent advances in photon counting detection technology have led to significant research interest in X-ray imaging. The photon counting detector has been characterized in different aspect including but not the least in energy resolution, detector sensitivity and many more by measuring its count rate performance, and its energy response. Recently the relationship between X-ray photocurrent response and photon counting performance of CdZnTe under different operational applied biases was investigated which showed the quadratic dependence between the experimental critical flux and applied bias. In other study a micoCT was designed with the integration of photon counting detector to investigate the spectroscopic performance of PCD by acquiring tomographic images of cylindrical PMMA phantom contains holes filled with different materials. This study showed that the multi energy threshold energy ranges allowed to measure target-to-background contrast at various energy ranges compared with EID detector which may useful to enhance the differentiation of various materials with different attenuation coefficient energy dependences. Another experiment-based study was developed to investigate the Photon counting spectroscopic CT (PCS_CT) imaging of anatomical soft tissue with clinically relevant size. This study showed several new findings including the effect of soft tissue non-uniformity on image artifacts, unique status of higher energy bin, and materials dependent visualization in spectroscopic image series [8-24,26-27].

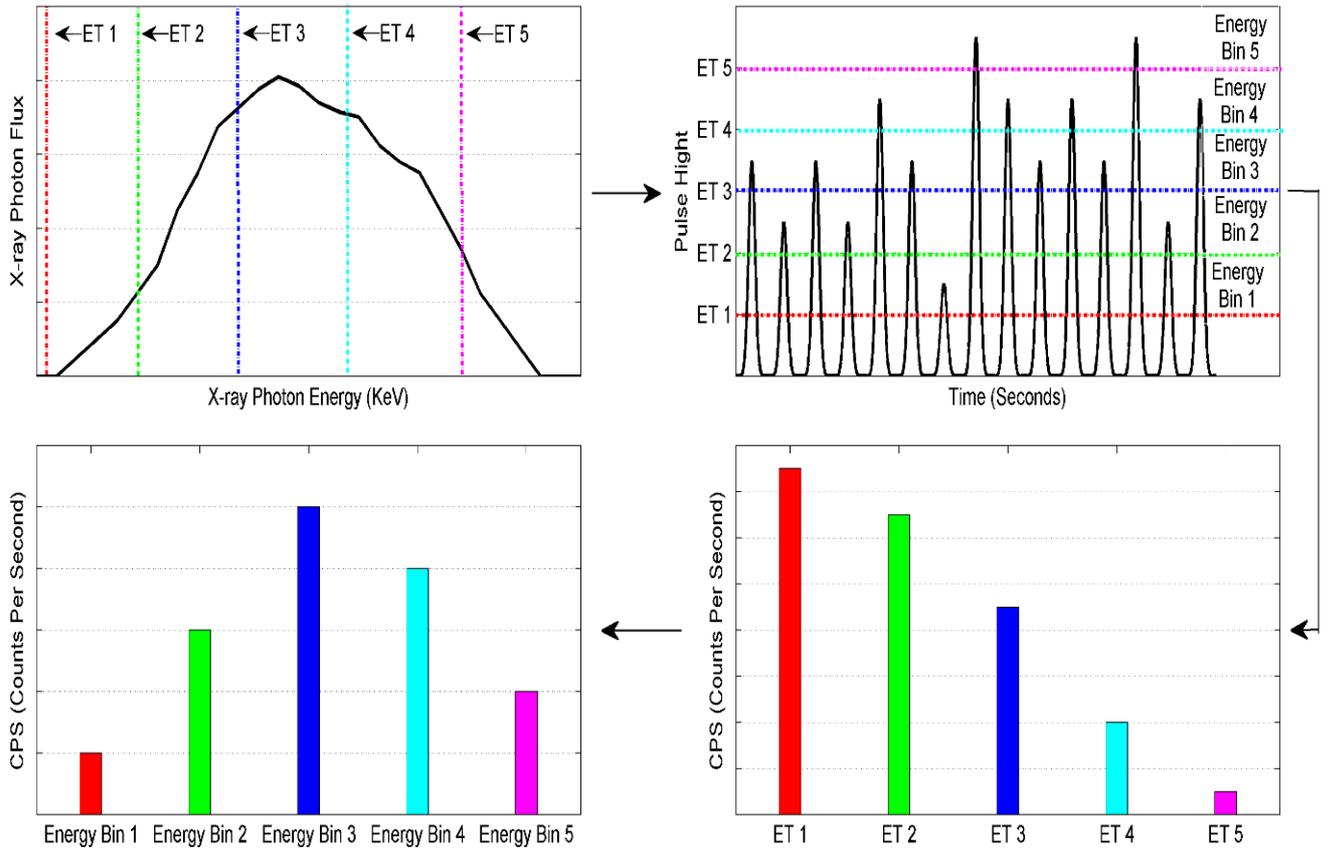

**Figure 1.** The X-ray photon detection using a PCD. (Top left) Five user-defined energy thresholds at a typical low kVp X-ray tube energy spectrum to form five different energy bins, (Top right) The pulse height generated by the X-ray photon is proportional to its energy, (bottom right) Integrated counts in each energy bin form a reverse S-curve and (bottom left) the total counts in each energy bin estimated by separating the reverse S-curve [2].

PCDs does not only allow energy-selective imaging with a single X-ray exposure, but they also provide lower noise and better contrast by offering more optimal weighting than EID [11]. These capabilities allow PCDs to be used in energy-selective photon counting X-ray computed tomography (CT) system for quantitative and efficient material differentiation and identification [12-15,26-27].

Thus, it is very important to evaluate the performance of the detector before using it for medical imaging application. The aim of this study was to characterize the MXA-128, a CZT-based linear array module (Imdetek, China), [19] detector 's linearity and its sensitivity in distinguishing various type of tissues in imaging.

## *2. Materials and Methods*

### *2.1. The MXA-128 CZT PCD*

The MXA-128 is a linear array detector module that consists of four main components including a high-voltage module, a linear array detector with a lead collimator, readout ASICs and a digital base board. These components were placed in a rectangular aluminum enclosure. The overall dimensions of the module were 295 x 210 x 70 mm3 (Figure 2).

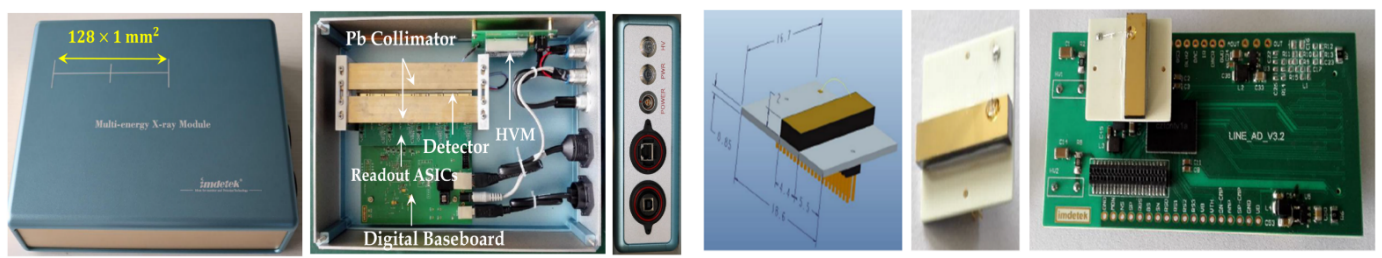

**Figure 2.** (Left) The size of Imdetek CZT linear array detector module (MXA-128), (middle) different components the linear array detector module and the main panel with high-voltage switch, power switch, power connector, GigE, and USB interface (top to bottom) , (Right) A schematic diagram of the crystal in mm measurement, (middle right ) An array of CZT PCD crystal, and crystal with readout ASICs.

### *2.1.1. High-voltage module*

The high-voltage module allows the detector's working voltage without external high voltage power supply. Thus, it is safe and convenient. The voltage ranges from 0 V to 1000 V and it can be adjusted by changing the value of the variable resistor connected to the half-voltage module [15].

### 2.1.2. Linear array detector with lead collimator

The MXA-128 which is consists of eight individual detector components. The detector dimension is 16.7 x 18.6 x 7.6 mm3 (Figure 3). It can operate at temperatures ranging from –10°C to 45°C. Table 1 provides the detail specification of the detector [15].

Table 1. Specification of each linear array detector

| Individual Detector Dimension | 16.7 x 18.6 x 7.6 mm$^3$ |
|---|---|
| Pin space (space between two pins for connection) | 0.8 mm |
| Pin height (pin for connection) | 4.75 mm |
| Base panel dimension | 16.7 x 18.6 x 0.85 mm$^3$ |
| Electrode material | Au |
| Crystal dimension | 16.7 x 4.4 x 2.0 mm$^3$ |
| Pixel dimension | 0.9 x 0.9 m$^2$ |
| Pixel space | 0.1 mm |

### 2.1.3. Readout ASICs

Integration of the electronics in custom application-specific integrated circuits (ASICs) is mandatory to read out multi-elements' semiconductor detector. Table 2 shows specification of the ASICs readout module [15]

**Table 2.** Specification of the ASICs readout module

| Criteria | Unit |
| --- | --- |
| Power consumption | 150 mW/module |
| Low voltage | DC: 2.5 to 5 V |
| High voltage | DC: – 450 to 600 V |
| Energy range | 20 to 160 keV |
| Peaking time | 100 nanoseconds |
| Pixel size | 1 mm |
| Counting rate | 1.0 Mcps (Million counts per second)/pixel |
| Sampling frame rate | < 2000 FPS |
| Energy Bins | 1 to 5 energy bins |
| Dimensions | 70 x 33.4 x 15 mm$^3$/module |

A maximum of 65,535 incoming photons can be counted in each frame with the use of 16-bit counter. The number of acquired frames per second (FPS) was set to a maximum of 2000. For any given FPS, if the number of counts is greater than 65535, the modulus of the count was recorded. For example, if the detected photon counts were 80000, thus, 14465 counts were recorded (80000 – 65535 =14465). The MXA-128 consists of four modules, 2 detectors per module and each detector consists of 32 channels. This study used two modules. Each channel allows the configuration of the channel parameters and data readout procedure through the serial mode. The X-ray photons in the energy range of 20 keV to160 keV was detected and read out by this module. The maximum energy detection capability can be increased up to 250 keV.

### *2.1.4. Digital base board*

The digital base board consists of a field programmable gate array and control panel for serial splicing [15]. It has the options for both USB- and GigE-based communication (Figure 2). However, only GigE was configured during the characterisation of the detector in this study because of GigE data transmission rate is fast as compared to the USB protocol.

### *2.1.5. Data acquisition (DA) interface*

It allows the user to define the FPS and the total number of frames to be acquired with a maximum of 16000 frames. The acquisition time is as shown in Equation (1).

$$\boldsymbol{Acquisition\ Duration} = \frac{\boldsymbol{Total\ Number\ of\ Frames}}{\boldsymbol{FPS}} \quad \textbf{(Eq. 1)}$$

The energy threshold also was set by the interface and by using an 8-bit value (0–255). The MXA-128 module allows the user to select a maximum of five different non-overlapping energy bins. Other PCD modules made by different manufacturers allow the user to set variable numbers of energy bins [17-19]. An 8-bit data value represents 0 to 255 was used to set the threshold to determine the energy width of each bin. The energy was related to the threshold as shown in Equation (2), which is specific for the PCD detector used in this study.

$$\boldsymbol{Energy} = (\boldsymbol{1.33 \times Threshold}) + \boldsymbol{4.698} \quad \textbf{(Eq. 2)}$$

This study suggested threshold setting for each of the five energy bins where Threshold 1 < Threshold 2 < Threshold 3 < Threshold 4 < Threshold 5, with the boundary conditions of Threshold 1 > 12 and Threshold 5 < 255, corresponding to minimum and maximum energies of 21 keV and 342 keV, respectively. If the threshold value was set to 255, it will disable the energy bin. The MXA-128 PCD also provides an option to work as a conventional EID by setting the first energy threshold to 15 and the remainder of the energy thresholds to 255.

## *2.2. Experimental setup*

The MXA-128, a CZT-based photon counting detector linear array module (Imdetek, China) was characterized for optimum FPS against detected X-ray photon energy, sensitivity to the X-ray tube current, sensitivity to the X-ray tube voltage and energy threshold and sensitivity in quantifying five different samples including fat, liver, muscle, contrast medium (Potassium iodide) and paraffin wax with different thickness. The experiments were conducted using a bench-top Phywe XRD system (Phywe, 2014) that includes a tungsten 35 W X-ray tube with a maximum continuous output of 35 kVp at 1.0 mA with a target angle of 190, The MXA-128 module with the Phywe XRD system is shown in Figure 5. The MXA-128 CZT detector was placed at 40 cm from the Phywe X-rays source for attaining the best performance as per the recommendation of the Phywe XRD system user manual.

Figure 5 shows the experimental system of an X-ray source and MXA-128 detector module. It comprises of a laboratory bench-top Phywe X-ray diffraction (XRD) system (Gottingen, Germany) [13] with a tungsten target, X-ray tube with a maximum continuous output of 35 kVp at 1.0 mA and a target angle of 19°, and the MXA-128 CZT detector module.

The optimum frame rate of the detector was set and five detector threshold energy bins (21 keV, 25 keV, 29 keV, 31 keV, and 45 keV) were used for image acquisition of the scanned object for experiment setup. The MXA detector module was internally attached to a DA interface system that was controlled by customized software (MXA) developed specifically for image acquisition of the scanned object.

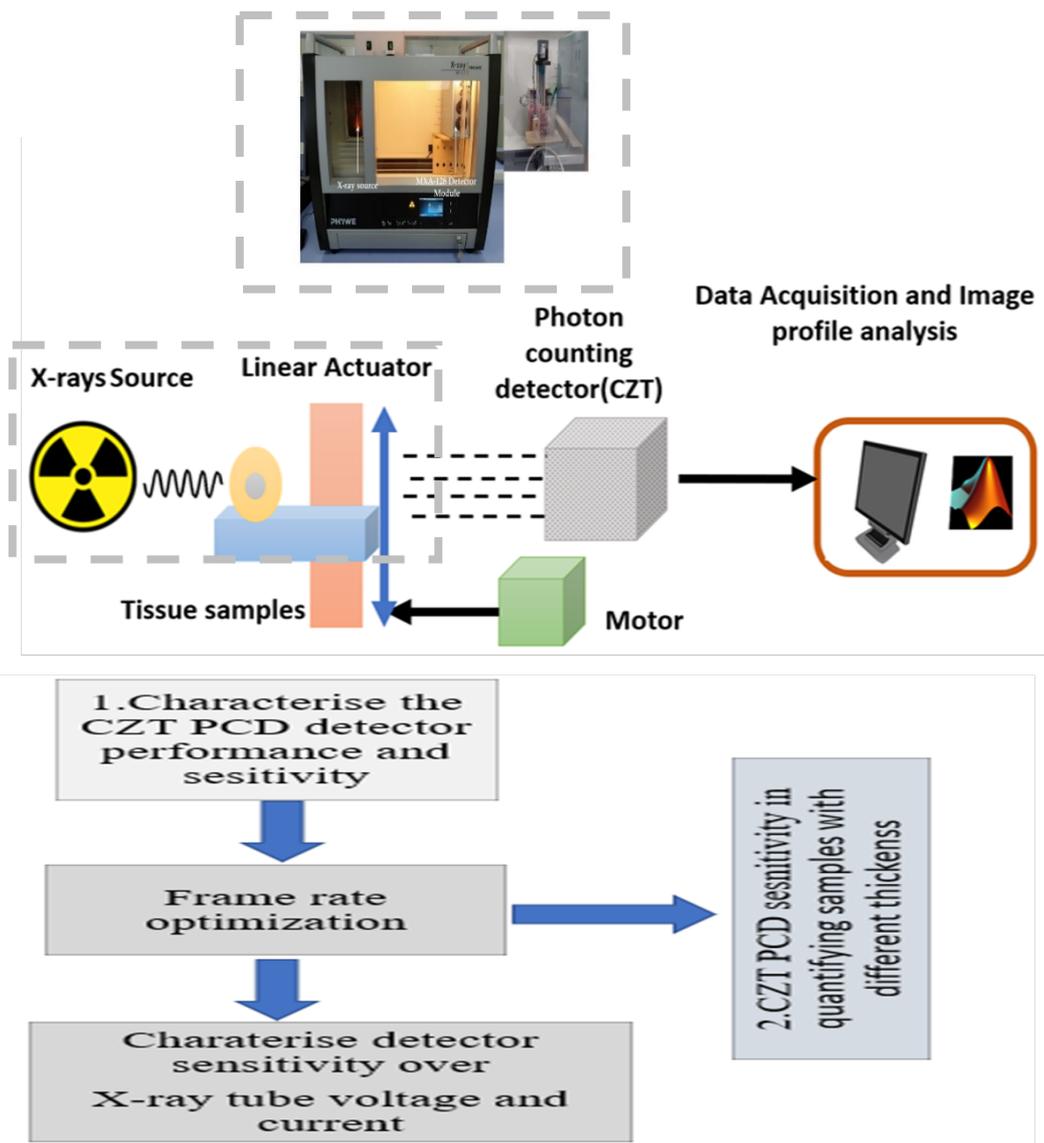

**Figure 3.** (Top) The experimental setup of the X-ray source in Phywe XRD system (Phywe, 2014). The X-ray tube was placed on the left, while the MXA-128 detector module was placed on the right of the system, (Bottom) Flowchart of Experiments setup and its procedure.

### 2.2.1 Frame rate optimisation

The voltage of the X-ray tube was set to 35 kVp with two different current settings (1.0 and 0.5 mA) to find the optimum FPS at various X-ray energies ranges (Bin), with different threshold values of PCD detector was set at 12, 15, 18, 20, 22 and 25 corresponding to 21, 25, 29, 3, 34 and 38 keV calculated using formula stated in Equation 2 which was used for both the current settings. The other bins were disabled by setting their threshold values to 255 according to the instruction of the manufacturer. For each threshold and current level, the FPS was first set at 100 and gradually reduced until the difference between two consecutive FPSs is minimal, and the total number of counts does not fall below 90% of the saturation limit of 65535 (i.e., approximately 59000). The lowest FPS between the two is then determined as the optimum FPS for that X-ray tube voltage, current and energy threshold.

### *2.2.2 Detector Sensitivity to X-ray tube Current*

The tube voltage was first set at 35 kVp and 25 kVp to characterise the sensitivity of the detector to X-ray tube current. The current was then varied for both voltages starting from 0.1 mA to 1.0 mA with increments of 0.1 mA. The energy threshold value of Bin 1 was set to 4 corresponding to 10 keV. The other bins were disabled by setting their threshold values to 255. The optimum FPS determined from the previous step for each voltage setting with 1.0 mA current was used to count the incoming X-ray photons.

### *2.2.3 Detector Sensitivity to X-ray tube Voltage and Energy Threshold*

For the characterisation of the detector sensitivity to the X-ray tube voltage and energy threshold, two separate scans were performed ranging from 10 to 38 keV with 10 non-uniform and non-overlapping energy threshold settings corresponding to 10 energy bins. The user defined threshold values and corresponding threshold energies (calculated using Equation 2) are shown in Table 3.

### *2.2.4 Detector Sensitivity in quantifying samples with different thickness.*

The x-ray transmission in CPS of the five different samples including fat, liver, muscle, contrast medium (potassium iodide), and paraffin wax) at six thicknesses ranging from 1 cm to 6 cm with steps of 1 cm were evaluated in this study. A unique stair-step chamber was fabricated from Plexiglass with a dimension of 21 cm x 21 cm and with various thicknesses to hold the samples for imaging purpose. The samples were placed in the stair-step sample chamber to obtain measurements for the six thicknesses (Figure 4).

The same phantom was also designed in vGATE Monte Carlo based simulation software to validate the physical results extracted in the sated experiment setup in section 2.2.4. The DA system of the detector module can easily measure the degradation intensities of photon counts based on sample thickness and the energy threshold during scanning process. The DA of the detector module was acquired for all five sample types and thicknesses. The intensity profile was stored in the .dat and .txt file formats. These files were imported into the MATLAB software environment, and image processing was performed to obtain the CPS values for each type of sample.
The measured CPS of all samples was obtained at various energy windows.

Figure 5 shows the X-ray transmission profiles obtained for fat sample over different thickness of fat sample, the X-ray transmission profiles for others samples i.e., liver, muscle and paraffin wax and contrast media were also plotted for the measurement of (CPS) counts. (Figure 3). Shows the complete experiment setup adopted for the study.

### *2.2.5 Simulation.*

A Monte Carlo based GATE simulation was also performed for the MXA-128 module equipped with a CZT PCD detector. The MXA-128 module equipped with a CZT PCD detector was modeled in GATE simulation. Pixel size for the detector was set to (0.4 *0.3 * 3.0 mm3) with 150 * 200 pixels in one module. We created five digitizer branches to create five energy bins (threshold windows) for acquiring images. The five digitizer branches were created for setting five energy bins (threshold windows) for obtaining image at these five energy windows. To define the same energy distribution as the physical source, the same histogram profile has been created in GATE simulation. The X-ray source was also created in GATE simulation to create the same histogram profile used to define the same energy distribution as in physical source. A geometrical configuration similar to the one used in a physical experiment set-up was used to simulate the phantom experiments. The phantom experiments were simulated using the same geometrical configuration used in a physical experiment set-up. We have also modeled a stair-step phantom chamber in solid works and imported it into GATE simulation for placing liver and muscle samples for scanning. A stair-step chamber was also modeled in solid works and imported it's in simulation for placing Liver and Muscle sample for scanning these sample with the integration of GATE simulation.

The simulation of X-ray projection results for different energy windows are provided in Figure -10. The X-ray transmission simulated results of the detector module were acquired for Liver and Muscle sample. The intensity profile was stored in .dat file formats. These files were than imported into the MATLAB software environment, and image processing was performed to obtain the CPS values for Liver and Muscle sample for validating the measured CPS of these samples with GATE simulation.

**Table 3:** Threshold values and corresponding threshold energies

| No. | Threshold Value | Threshold Energy |
|---|---|---|
| 1 | 4 | 10 |
| 2 | 6 | 13 |
| 3 | 8 | 15 |
| 4 | 10 | 18 |
| 5 | 12 | 21 |
| 6 | 15 | 25 |
| 7 | 18 | 29 |
| 8 | 20 | 31 |
| 9 | 22 | 34 |
| 10 | 25 | 38 |

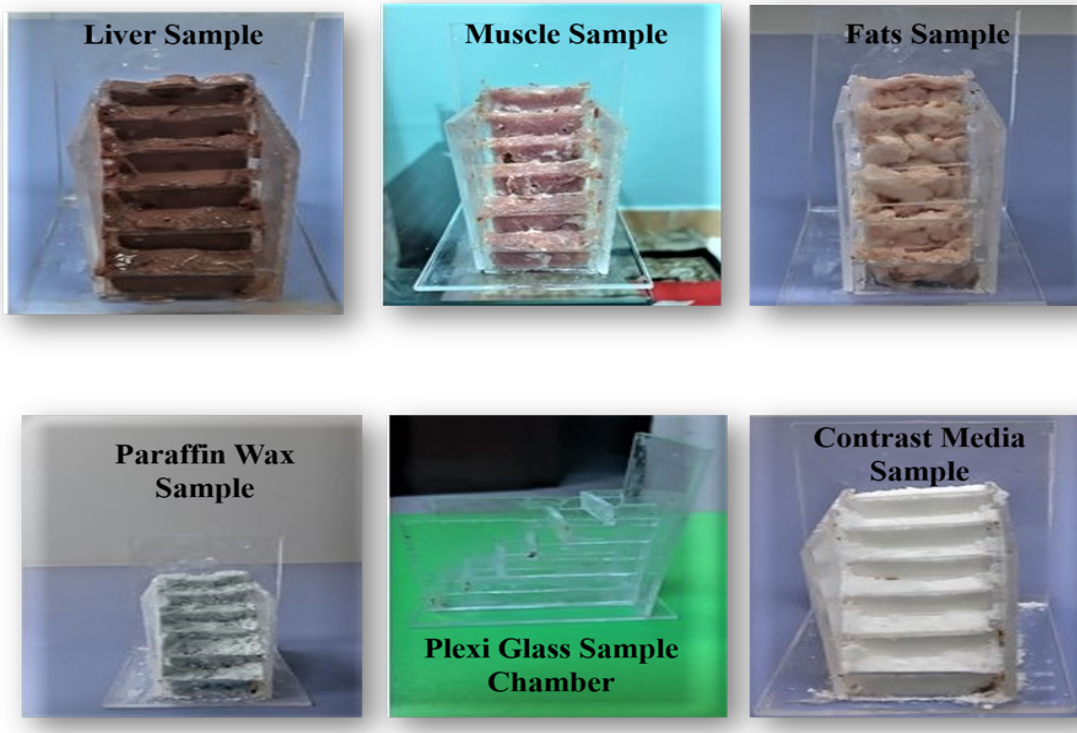

**Figure 4.** A unique stair-step chamber was fabricated from Plexiglass with a dimension of 21 cm x 21 cm to hold the fat, liver, muscle, paraffin wax and contrast media at six different thickness.

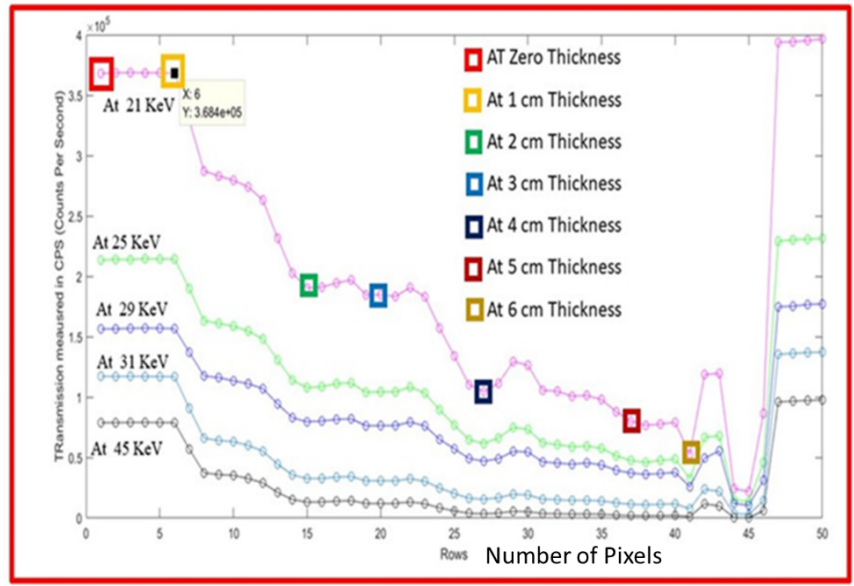

**Figure 5.** The X-ray transmission profile obtained from fats sample plotted in MATLAB software for six different thickness.

## 3. Results and Discussion

3.1. Frame rate optimization

Figure 6 (left) shows the relationship between the optimum FPS, X-ray tube current and detected X-ray energy. For a 35 kVp X-ray tube voltage and 1.0 mA current, the peak of X-ray photon flux is expected between 21 to 23 keV of the detected photon energy. An optimum FPS was found at this energy level in this study 12The optimum FPS at this energy level is 12, because12,at, This was the setting where the scanning phantom produced the best image quality and the total number of CPS does not fall below 90% of the saturation limit of 65535 (i.e., approximately 59000) of the CZT detector used in this study.

The flux of incoming X-ray photon reduces as the energy of the X-ray photons increases at 35 kVp tube voltage. This is due to optimum FPS reduces with the increase of the X-ray photon energies at 35 kVp. The optimum FPSs at 21, 25, 29, 31, 34 and 38 keV are 12, 10, 6, 4, 4 and 2 respectively for 1.0 mA tube current as shown in Figure 6 (left). For the same voltage of 35 kVp, when the current is reduced (e.g., 0.5 mA), the optimum FPS is also reduced.

Since the detector performance is not always reliable below the energy level of 20 keV due to electronic noise as recommended by manufacturer, the optimum FPS below 20 keV was not determined.

### *3.2. Detector Sensitivity to X-ray tube Current*

With the increase of the X-ray tube current, the maximum CPS of the detector also increases as shown in Figure 6 (right) and they are linearly related. The maximum CPS for 35 kVp are higher than that of 25kVp for each individual current level. However, the linear relationships are present for the latter case as well. The rate of change of maximum CPS (slope) for 35 kVp is higher than 25 kVp and the linear fitted lines for both the settings are given by which were calculated during Linear regression fitted slope value.

$$\text{Maximum CPS35} - \text{kVp} = [362880 \times \text{X-ray Tube Current (mA)}] + 37403 \quad \textbf{(Eq. 3)}$$

$$\text{Maximum CPS25} - \text{kVp} = [83530 \times \text{X-ray Tube Current (mA)}] + 67239 \quad \textbf{(Eq. 4)}$$

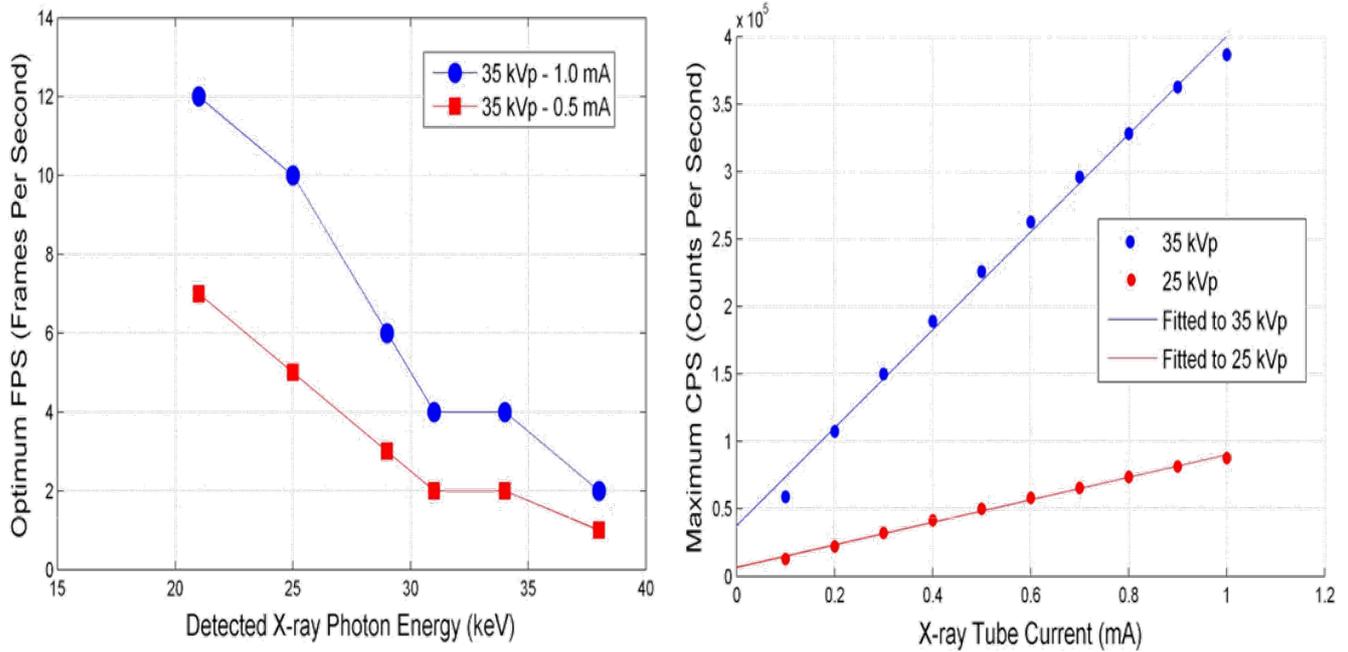

**Figure 6:** Left: relationship between optimum FPS, X-ray tube current and detected X-ray energy. Right: relationship of maximum counts per seconds (CPS) with X-ray tube current.

## *3.3. Count rate performance against X-ray tube Voltage and Detector Energy Threshold*

The maximum CPS provided by the CZT detector increases with either the X-ray tube voltage or current. No change in the detected X-ray energy spectrum was observed for 35 kVp though the current were varied by a factor of 2. A similar observation is made for 25 kVp as well. The SD of 8 measurements is negligible indicating that the detector is robust against measurement variations. Since the tungsten characteristic X-ray Kα lines (57–59 keV) are only visible for tube voltages above 90 kVp, at 25 and 35 kVp the characteristics X-ray is not visible. Since below 20 keV the performance of the detector was not reliable due to the presence of electronic noise, the detected X-ray photon spectrum does not follow the ideal X-ray spectrum for different tube voltage and current settings.

### *3.4. CZT PCD sensitivity in quantifying samples with different thicknesses*

The CZT PCD was used to scan various types of samples at five threshold energies ranging from 21 keV to 45 keV at a rate of 12 FPS. The CPS decreased with increasing sample thickness. This is due to higher attenuation values were detected in thicker samples and at lower threshold energies. As the energy level increases, the relationship between CPS and sample thickness became linear gradually due to lower attenuation at higher energy values. CPS of the sample also decreases with the increasing photon energy (Figure 7). The multiple measurement in this study proved that the CZT PCD performance was consistent across the energies at different thickness of samples (Figure 8). In addition, Figure 7 shows the transmission of X-ray profile of each sample scanned in this study at various energies and at different thickness, while the Figure 8 presents the transmission of X-ray profile for each sample at various energies obtained at five repeated measurements These figures show the consistency of detector response at all these five measurements at different energies as shown in **Figures 7–8**.

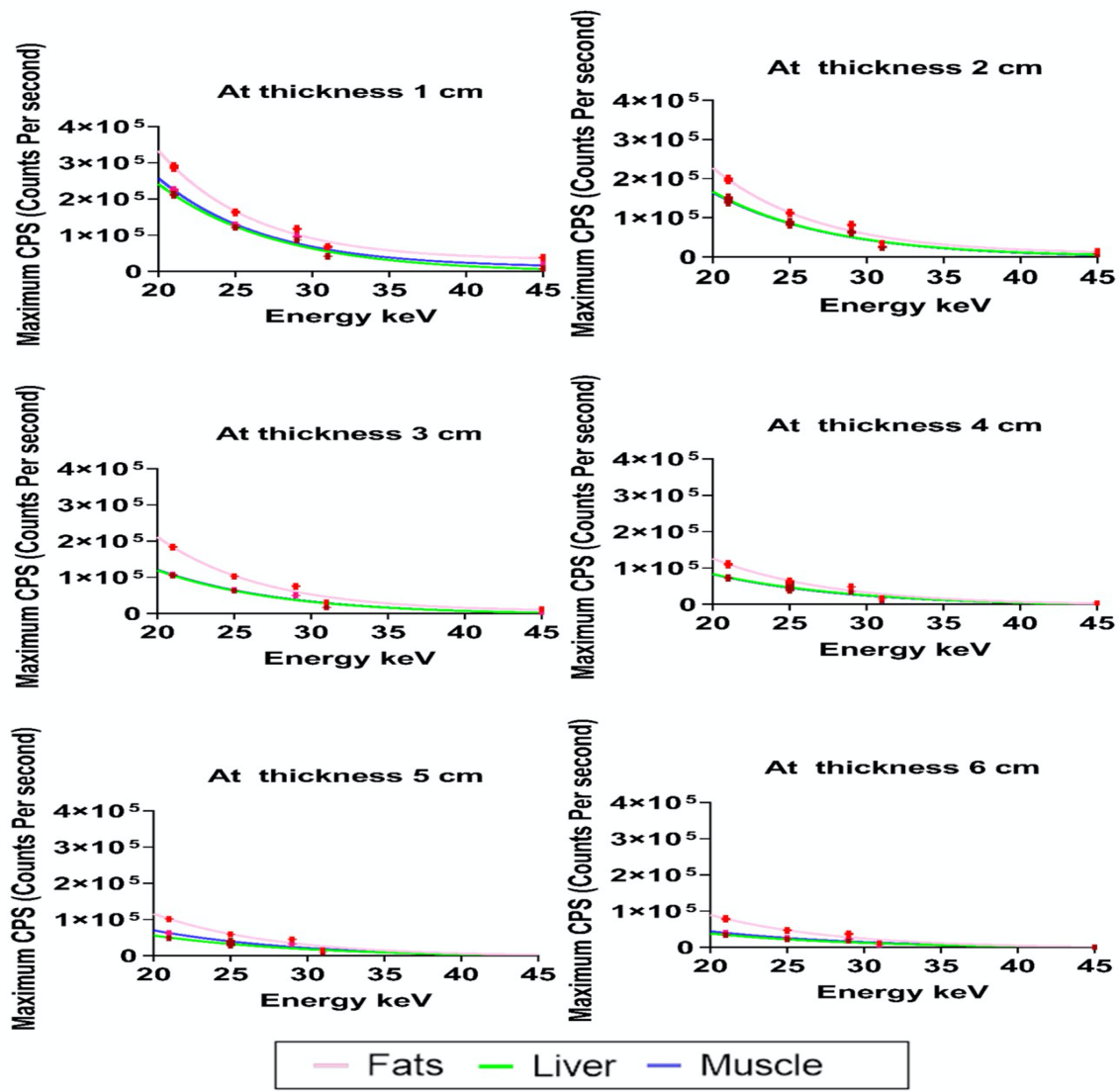

**Figure 7.** The transmission of X-ray photons profile measured in count per second for different samples including fat, liver, muscle, contrast media and paraffin wax scanned with a CZT detector at five X-ray energy thresholds ranging from 21 keV to 45 keV at various sample thickness ranging from 1 cm to 6 cm, with a step size of 1 cm.

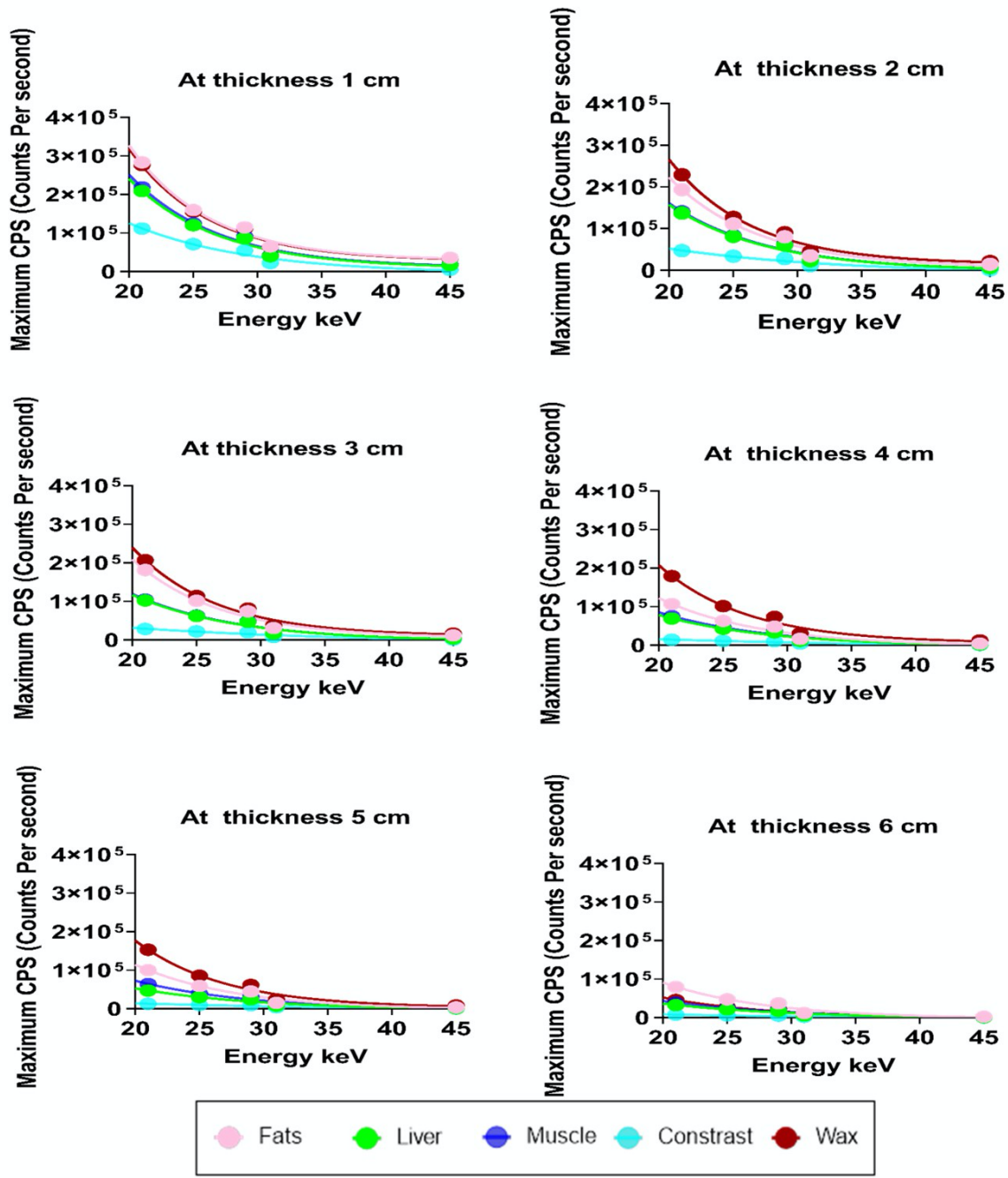

**Figure 8.** The X-ray transmission measured in count per seconds of five data sets with known thickness samples including fats, liver, muscle, contrast media and paraffin wax scanned with a CZT detector at five X-rays energies threshold ranging from 21 keV to 45 keV, with five measurements which were illustrated by the mean and standard deviation plotted in the figure. The red color represents the error bar of five data set of all sample with known thickness. Small standard deviation indicates that the detector provides accurate measurement with very small deviation.

### *3.4.1. Gate X-ray transmission simulation result for validation*

The X-ray transmission profile passing through liver and muscle sample was simulated with same configuration as physical experiment performed in the laboratory. Results from the x-ray simulation projections are shown in Figure 9 for liver samples. The X-ray projection images as shown in Figure 9 and its enegry versus graphs for liver and muscle samples as shown in Figure 10-11, at five different energy bins ( 25 keV to 45 keV), indicate that the projected transmission for liver and muscle samples decreases with increasing thickness of the samples, which were placed in the stair step multi thickness phantom chamber, and increases for high energy samples. The results show that transmission decreases between 31 keV and 45 keV because the X-ray tube has a maximum flux at 23 keV. Despite this, The transmission versus thickness becomes linear at all higher energy detector bins due to lower transmission detected by the detector [25, 28] . The p-values were also calculated between the measured sample data of muscle , liver respectvely with their simulated CPS values at each phantom thickness and energy bins respectively as shown in Table 4 , for this Wilcoxon rank sum test was used to test whether two populations have the same continuous distribution of data or not. The p-values caluclated for each phantom sample thickness at each energy bins were found P > 0.05 , which means tis study accepted the null hypothesis which mean that the difference between two populations is not significan which validated this stduy too.

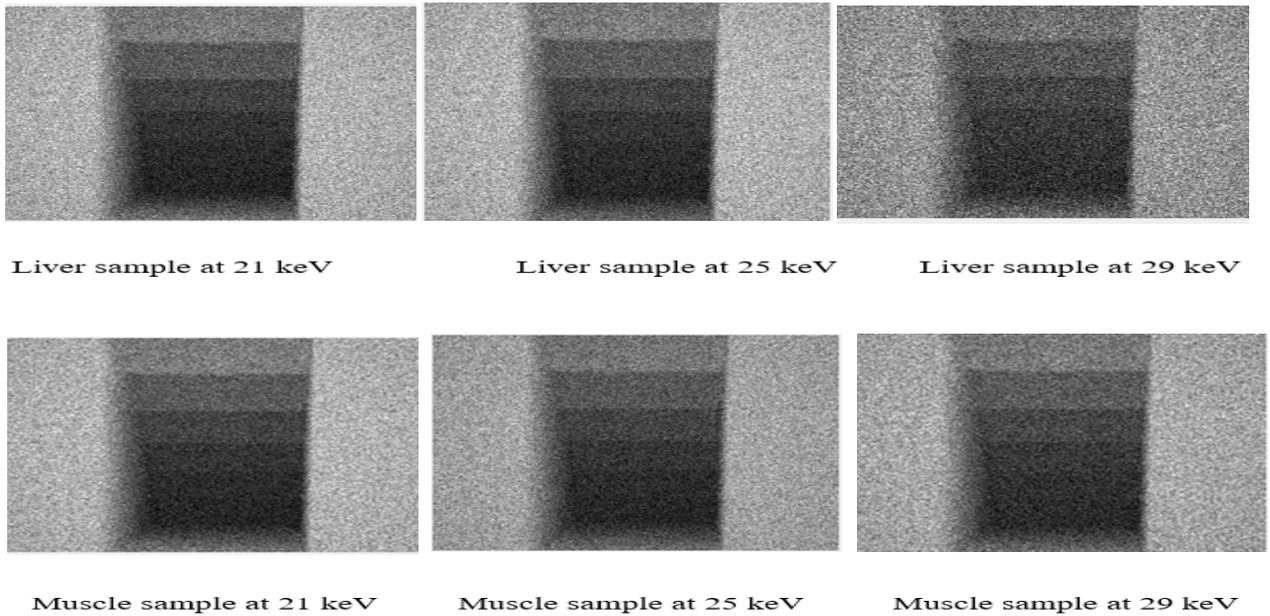

**Figure 9.** At Top: Liver phantom image acquired at 21 , 25 and 29 keV CZT detector threshold energy , At Bottom: Muscle phantom image acquired at 21 , 25 and 29 keV CZT detector threshold energyrespectively.

Figures 10-11. shows the comparative graph between the transmission of X-ray photons profile measured in count per second for different Liver and muscle sample respectively scanned with a CZT detector modeled in GATE simulation for validating the measured results at five X-ray energy thresholds ranging from 21 keV to 45 keV which almost giving the same trends as compared to the CPS value detected by the physical detector experiment setup.

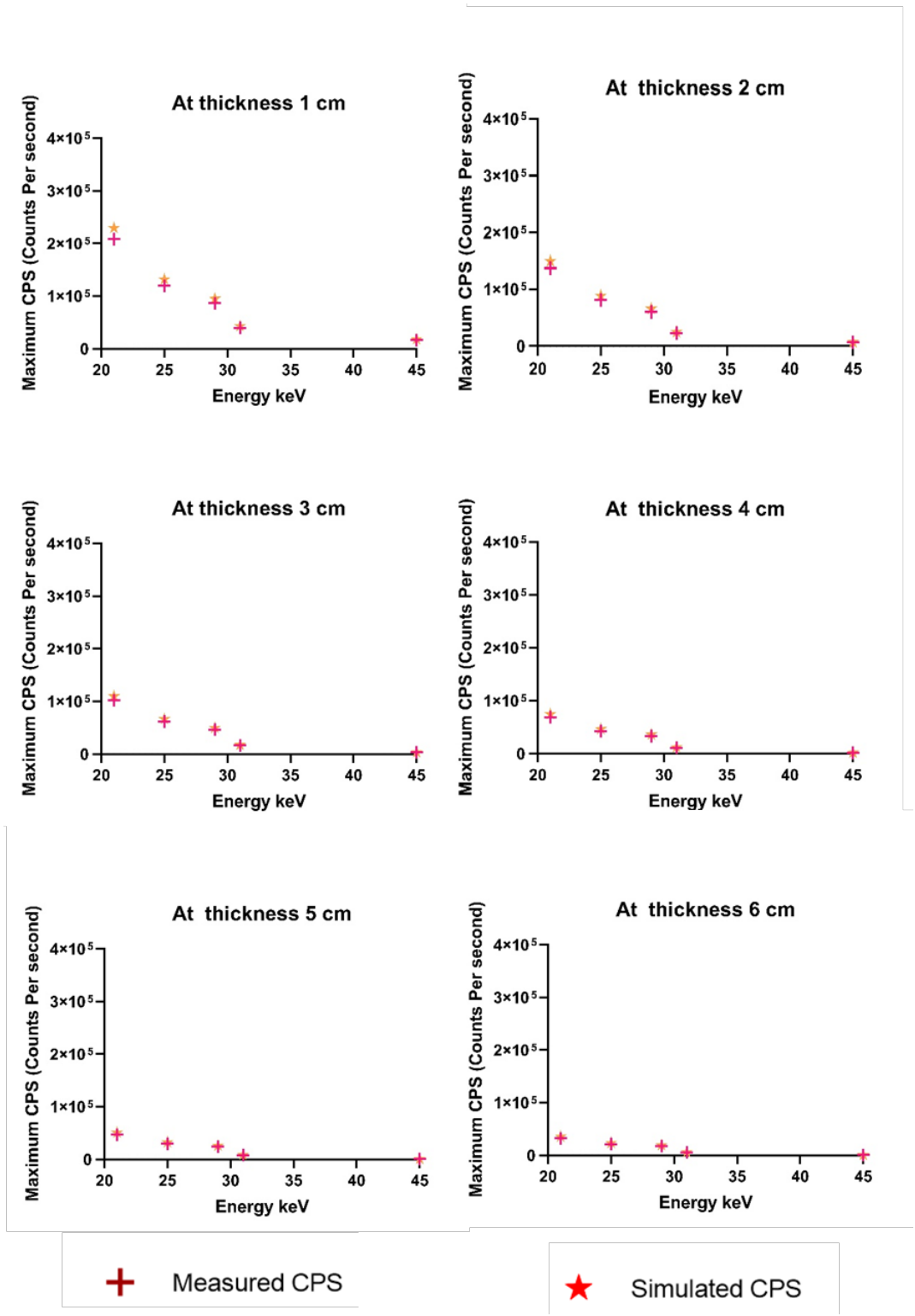

**Figure 10.** The transmission of X-ray photons profile measured in count per second for different Liver sample scanned with a CZT detector modeled in GATE simulation for validating the measured results at five X-ray energy thresholds ranging from 21 keV to 45 keV.

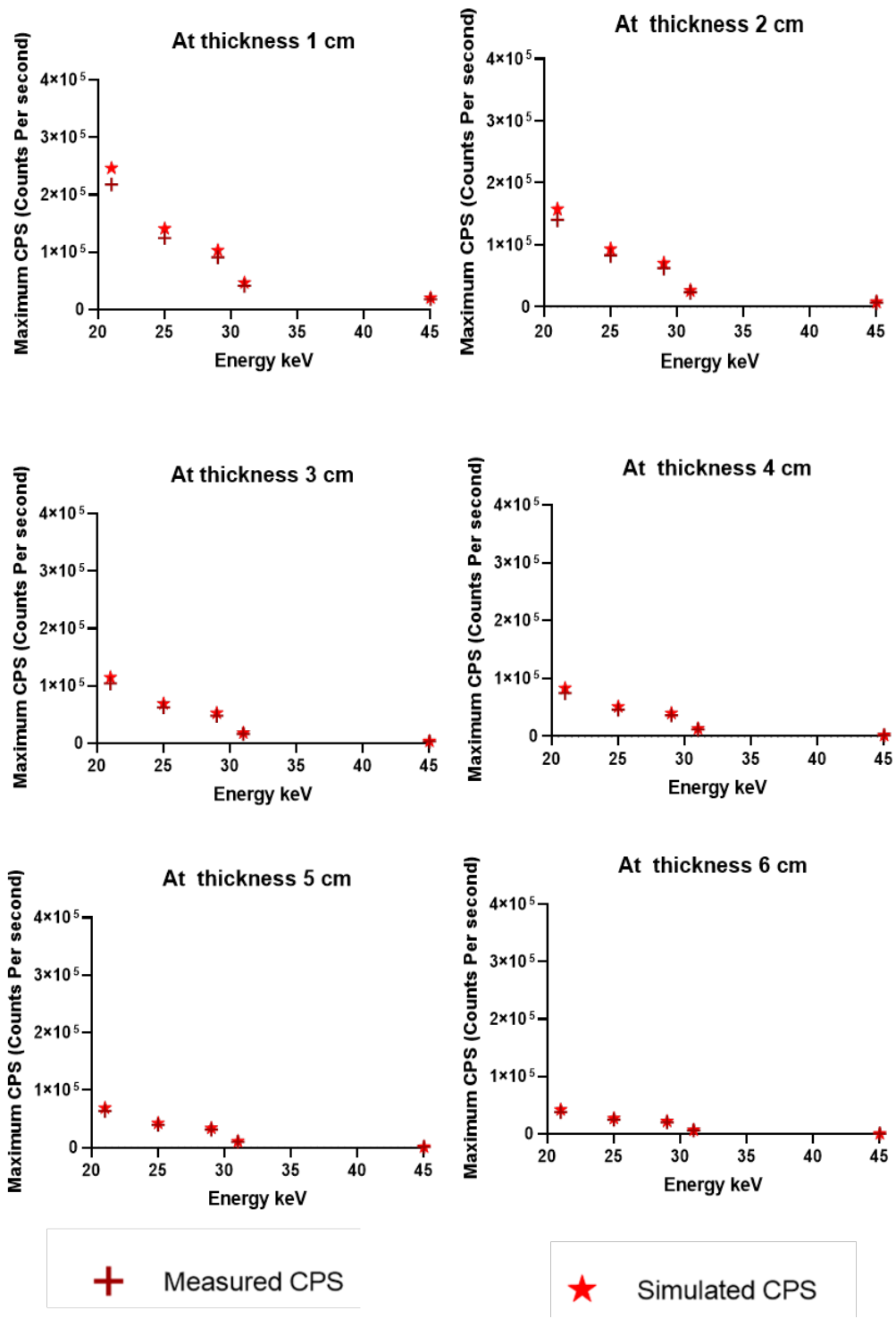

**Figure 11.** The transmission of X-ray photons profile measured in count per second for different muscle samples scanned with a CZT detector modeled in GATE simulation for validating the measured results at five X-ray energy thresholds ranging from 21 keV to 45 keV.

**Table 4:** p-values between measured and simulated sample data for muscles and liver sample

| | Thickness (cm) | | | | | | | | | | | | |
|---|---|---|---|---|---|---|---|---|---|---|---|---|---|
| | 1 | | 2 | | 3 | | 4 | | 5 | | 6 | | |
| | M | L | M | L | M | L | M | L | M | L | M | L | Energy Bins(keV) |
| P-value | 0.78 | 0.86 | 0.83 | 0.87 | 0.86 | 0.89 | 0.84 | 0.87 | 0.89 | 0.88 | 0.86 | 0.84 | 21 |
| | 0.82 | 0.93 | 0.83 | 0.87 | 0.86 | 0.89 | 0.84 | 0.87 | 0.89 | 0.88 | 0.86 | 0.84 | 25 |
| | 0.95 | 0.47 | 0.83 | 0.87 | 0.86 | 0.89 | 0.84 | 0.87 | 0.89 | 0.88 | 0.86 | 0.84 | 29 |
| | 0.82 | 0.88 | 0.83 | 0.87 | 0.86 | 0.89 | 0.84 | 0.87 | 0.89 | 0.88 | 0.86 | 0.84 | 31 |
| | 0.76 | 0.83 | 0.83 | 0.87 | 0.86 | 0.89 | 0.84 | 0.87 | 0.89 | 0.88 | 0.86 | 0.84 | 45 |

*Muscle = M, Liver = L

## *4. Conclusion*

We have evaluated the performance of the MXA-128 module equipped with a CZT PCD that measured the X-ray photons with five energy bins, and also a number of samples (phantoms) were used to evaluate the physical performance and imaging quality of the detector, A study calculated 12 FPS as an optimal frame rate that follows the X-ray source spectrum used in this study, Moreover, the study showed that the CPS has a linear relationship with the current in the X-ray tube. Furthermore, it was noted that tissue sample thickness also affected the X-ray transmission which was calculated as CPS. The results of the experiments showed that the photon counting rate was linear up to a count rate of $0.4 \times 10^6$ counts/sec/pixel. The maximum X-ray transmission (CPS) was counted $2.8330 \times 10^5$ for fats sample which followed by paraffin wax of $2.7610 \times 10^5$, $2.1800 \times 10^5$ for muscle sample, $2..0860 \times 10^5$ for liver sample and $1.1100 \times 10^5$ for X-ray contrast medium respectively at 21 keV X-ray energy bin (i.e., At lower detector threshold energy setting), at 1 cm of phantom thickness while the minimum X-ray transmission ( CPS) was counted 135 for X-ray contrast medium sample followed by liver sample of 423, 561 for muscle sample, 785 for paraffin wax sample and 2042 for fats sample respectively at 45 keV X-ray energy bin (i.e., At higher detector threshold energy setting), at 6 cm of phantom thickness was set. The Monte Carlo based GATE simulation was also performed for validating the experiment results trends for muscle and liver sample at various phantom thickness. The simulation results showed that the sensitivity of the detector was considerably uniform compared with experimental results along with their p-values were also calculated which shows that there is no statically Signiant large difference between the measured CPS versus simulated CPS for both muscle and liver respectively. Calculated between these measured The high sensitivity and linearity of the detector which was noticed between 87 to 90 % in this study, which made PCD detector suitable to be used in X-ray based medical applications as well as for preclinical applications. The high sensitivity and linearity of the detector make it suitable to be used in X-ray based medical applications (e.g., radiography, CT, mammography etc.) as well as for preclinical applications where the object is small (e.g., mouse). However, to obtain optimum image quality and valuable diagnostic information in all of the available five energy spectrums, the threshold and width of each energy bin needs to be optimized for a specific application based on the size and density of the object.

## Acknowledgment

This study was funded by the Deanship of Scientific Research (DSR), Imam Abdulrahman Bin Faisal University, Dammam, Saudi Arabia (grant number: 2018-033-Eng).

Conflicts of Interest: The authors declare no conflict of interest.

## Conflict of Interest

The authors have no conflict of interests in this study.

## List of References

[1]     Wang X, Meier D, Mikkelsen S, Maehlum GE, Wagenaar DJ, Tsui BMW, et al. Micro CT with energy-resolved photon-counting detectors . Phys Med Biol 2011;56:2791–816. https://doi.org/10.1088/0031-9155/56/9/011 .

[2]     Taguchi K, Iwanczyk JS. Vision 20/20: Single photon counting X-ray detectors in medical imaging: Vision 20/20: Photon counting detectors. Med Phys 2013;40:100901. https://doi.org/10.1118/1.4820371 .

[3]     Wang X, Meier D, Taguchi K, Wagenaar DJ, Patt BE, Frey EC. Material separation in X-ray CT with energy resolved photon-counting detectors. Med Phys 2011;38:1534–46. https://doi.org/10.1118/1.3553401 .

[4]     Shikhaliev PM. Energy-resolved computed tomography: first experimental results. Phys Med Biol 2008;53:5595–613. https://doi.org/10.1088/0031-9155/53/20/002 .

[5]     Ding H, Ducote JL, Molloi S. Breast composition measurement with a cadmium-zinc-telluride based spectral computed tomography system . Med Phys 2012;39:1289–97. https://doi.org/10.1118/1.3681273 .

[6]     Schlomka JP, Roessl E, Dorscheid R, Dill S, Martens G, Istel T, et al. Experimental feasibility of multi-energy photon-counting K-edge imaging in pre-clinical computed tomography .Phys Med Biol 2008;53:4031–47. https://doi.org/10.1088/0031-9155/53/15/002 .

[7]     Ren L, Zheng B, Liu H. Tutorial on X-ray photon counting detector characterization . J X-Ray Sci Technol  2018;26:1–28. https://doi.org/10.3233/XST-16210 .

[8]     Bornefalk H, Danielsson M. Photon-counting spectral computed tomography using silicon strip detectors: a feasibility study . Phys Med Biol 2010;55:1999–2022. https://doi.org/10.1088/0031-9155/55/7/014 .

[9]     Cho H-M, Barber WC, Ding H, Iwanczyk JS, Molloi S. Characteristic performance evaluation of a photon counting Si strip detector for low dose spectral breast CT imaging: Characteristic evaluation of a photon counting Si strip detector . Med Phys  2014;41:91903. https://doi.org/10.1118/1.4892174 .

[10]     Y. Li et al, "Investigation on X-Ray Photocurrent Response of CdZnTe Photon Counting Detectors," Sensors (Basel, Switzerland), vol. 20, (2), pp. 383, 2020.

[11]     X. Wang et al, "MicroCT with energy-resolved photon-counting detectors," Physics in Medicine & Biology, vol. 56, (9), pp. 2791-2816, 2011.

[11]     S. Leng et al, "Photon-counting Detector CT: System Design and Clinical Applications of an Emerging Technology," Radiographics, vol. 39, (3), pp. 729-743, 2019

[12]     L. Ren, B. Zheng and H. Liu, "Tutorial on X-ray photon counting detector characterization," Journal of X-Ray Science and Technology, vol. 26, (1), pp. 1-28, 2018.

[13]     P. M. Shikhaliev, "Soft tissue imaging with photon counting spectroscopic CT," Physics in Medicine & Biology, vol. 60, (6), pp. 2453-2474, 2015.

[14]     Hsieh SS, Rajbhandary PL, Pelc NJ. Spectral resolution and high‐flux capability tradeoffs in CdTe detectors for clinical CT . Med Phys  2018;45:1433–43. https://doi.org/10.1002/mp.12799 .

[15]     Phywe xre 4.0 X-ray expert set unit 2015. https://www.phywe.com/en/xr-4-0-expert-unit-X-ray-unit-35-kv.html.

[16]     Cormode DP, Roessl E, Thran A, Skajaa T, Gordon RE, Schlomka J-P, et al. Atherosclerotic plaque composition: analysis with multicolor CT and targeted gold nanoparticles . Radiol  2010;256:774–82. https://doi.org/10.1148/radiol.10092473 .

[17]     Tremblay J, Bedwani S, Bouchard H. A theoretical comparison of tissue parameter extraction methods for dual energy computed tomography . Med Phys  2014;41:081905-n/a. https://doi.org/10.1118/1.4886055 .

[18]     Feuerlein S, Klass O, Hoffmann MH, Schlomka JP. Multienergy Photon-counting K-edge Imaging: Potenzial für eine verbesserte Gefäßdarstellung . RöFo - Fortschritte Auf Dem Gebiet Der Röntgenstrahlen Und Der Bildgeb Verfahren  2009;181. https://doi.org/10.1055/s-0029-1221318 .

[19]     MXA-128 X-ray Photon Counting Imaging System n.d. http://www.imdetek.com

[20]     Mousa, A., Kusminarto, K., & Suparta GB. A new simple method to measure the X-ray linear attenuation coefficients of materials using micro-digital radiography machine. Int J Appl Eng Res 2017;12:10589–10594.

[21]     Niu S, Zhang Y, Zhong Y, Liu G, Lu S, Zhang X, et al. Iterative reconstruction for photon-counting CT using prior image constrained total generalized variation . Comput Biol Med 2018;103:167–82. https://doi.org/10.1016/j.compbiomed.2018.10.022 .

[22]     Shikhaliev PM. Computed tomography with energy-resolved detection: a feasibility study . Phys Med Biol  2008;53:1475–95. https://doi.org/10.1088/0031-9155/53/5/020 .

[23]	Chitralekha, Kerur BR, Lagare MT, Nathuram R, Sharma DN. Mass attenuation coefficients of saccharides for low-energy X-rays . Radiat Phys Chem (Oxford, Engl 1993) 2005;72:1–5. https://doi.org/10.1016/j.radphyschem.2004.03.007 .

[24]	Smale LF, Chantler CT, de Jonge MD, Barnea Z, Tran CQ. Analysis of X-ray absorption fine structure using absolute X-ray mass attenuation coefficients: Application to molybdenum . Radiat Phys Chem (Oxford, Engl 1993) 2006;75:1559–63. https://doi.org/10.1016/j.radphyschem.2005.07.016 .

[25] K. Hameed et al, "Characterization of CdZnTe detector for spectral computed tomography," in 2019.

[26] M. Holbrook et al, "Development of a spectral photon-counting micro-CT system with a translate-rotate geometry," in 2018, . DOI: 10.1117/12.2293373.

[27] N. Kimoto et al, "A novel algorithm for extracting soft-tissue and bone images measured using a photon-counting type X-ray imaging detector with the help of effective atomic number analysis," Applied Radiation and Isotopes, vol. 176, pp. 109822, 2021.

[28]	A. Boke, "Linear attenuation coefficients of tissues from 1 keV to 150 keV," Radiation Physics and Chemistry (Oxford, England : 1993), vol. 102, pp. 49-59, 2014.